\documentstyle[aps,prl,epsf,twocolumn,floats]{revtex}
\begin{document}
\draft

\twocolumn[\hsize\textwidth\columnwidth\hsize\csname @twocolumnfalse\endcsname

\title{Measuring the Out-of-Equilibrium Splitting of the Kondo Resonance}
\bigskip
\author{Eran Lebanon and Avraham Schiller}
\address{ Racah Institute of Physics, The Hebrew University,
          Jerusalem 91904, Israel }
\date{\today}
\maketitle

\begin{abstract}
An experiment is proposed to measure the out-of-equilibrium splitting of
the Kondo resonance in an ultrasmall quantum dot, by adding a third,
weakly coupled lead to the standard two-lead quantum-dot system, and
sweeping the chemical potential of that lead. Fixing the voltage bias
between the source and drain leads, we show that the differential
conductance for the current through the third lead traces the
out-of-equilibrium dot density of states (DOS) for the two-lead
system. This enables one to measure the dot DOS in the presence
of an applied voltage bias. We show that this method is robust, and
extends also to the case where the coupling to the third lead is no
longer weak.
\end{abstract}

\bigskip

]
\narrowtext

The understanding of strong electronic correlations far from
thermal equilibrium is one of the challenging problems in
contemporary mesoscopic physics. A primary example is the
out-of-equilibrium Kondo effect, recently measured in ultrasmall
quantum dots~\cite{QD1,QD2,QD3,QD4}. In the Kondo effect,
an impurity moment undergoes a many-body screening by the
conduction electrons, producing a narrow low-energy resonance
in the impurity density of states (DOS) at low
temperatures~\cite{Hewson_book}. This Kondo resonance
is responsible for the well-known enhancement of the
low-temperature resistivity in dilute magnetic alloys, and
for the enhancement of the low-temperature conductance in
tunneling through small quantum dots with an odd number
of electrons. One of the striking predictions for the
out-of-equilibrium Kondo effect in quantum dots is the
splitting of the Kondo resonance in the dot DOS for
a finite applied bias~\cite{WM94}.

Although well-established theoretically, it remains unclear
whether one can measure the above splitting in the dot DOS.
The splitting of the Kondo resonance does not show up in the
differential conductance for the current through the dot
since, unlike in conventional noninteracting systems, the
DOS itself is strongly voltage dependent. Photoemission, besides
being limited in resolution, measures the ``occupied'' DOS, i.e.,
the product of the DOS and the effective distribution function.
Since the latter distribution is unknown away from thermal
equilibrium, one can not reliably extract the underlying DOS
using photoemission. A key assumption in conventional tunneling
measurements of the DOS is that the weakly coupled sample
and probe are each effectively in equilibrium. Clearly this
assumption breaks down for the quantum dot, which raises the
fundamental question: Can one actually measure the DOS of an
interacting mesoscopic system far from thermal equilibrium?

\begin{figure}[tb]
\centerline{ \vbox{ \epsfxsize=50mm \epsfbox {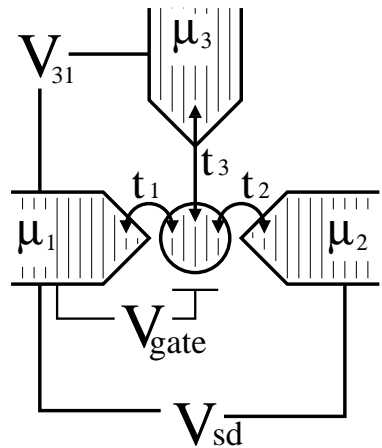}}
}
\vspace{10pt}
\caption{
    A schematic sketch of the proposed apparatus. An ultrasmall
    quantum dot is coupled by tunneling to three metallic leads,
    each of which is kept at a separate chemical potential. The
    corresponding tunneling matrix elements are controlled by
    varying the potential barriers, while the dot energy level
    is adjusted by applying a gate voltage. The dot level and
    the couplings are to be tuned such that the dot is
    in the Kondo regime, and the coupling to the third lead is
    much weaker than to the other two leads. Fixing the
    source-drain voltage bias at $\mu_2 - \mu_1 = eV_{sd}$ and
    sweeping the chemical potential $\mu_3$ by varying $V_{31}$,
    one measures the current $I_3(V_{31})$ between the dot and
    the third lead. Up to thermal broadening and rescaling, the
    differential conductance $G_3 = dI_3/dV_{31}$ traces the
    out-of-equilibrium two-lead dot DOS.
}
\label{fig:fig1}
\end{figure}

In this paper we show that the out-of-equilibrium splitting
of the Kondo resonance can indeed be measured, by adding a
third, weakly coupled lead to the standard two-lead quantum-dot
system~\cite{Goldhaber-Gordon_Pecs}.
Fixing the voltage bias between the source and
drain leads and sweeping the chemical potential of the
third lead, we show that the differential conductance
for the current through that lead traces the
out-of-equilibrium two-lead dot DOS, much in the same way as in
conventional tunneling measurements of the equilibrium DOS.
In this manner, one can measure the dot DOS in the presence
of an applied voltage bias. Furthermore, we find that the
splitting of the Kondo resonance can be observed also at
temperatures above the Kondo temperature, and that the
same qualitative features are obtained when the coupling
to the third lead is no longer weak. Contrary to a
previous proposal to probe the splitting of the Kondo
resonance using two capacitively coupled quantum
dots~\cite{Pohjola_etal}, our approach allows for
a direct measurement of the nonequilibrium DOS.

Figure~\ref{fig:fig1} shows the proposed experimental setting.
It consists of an ultrasmall quantum dot, coupled to three
metallic leads. Each lead is kept at a separate chemical 
potential $\mu_{\alpha}$ ($\alpha =1,2,3$), and is coupled
to the dot via a tunneling matrix element $t_{\alpha}$. The
energy level on the dot, $E_D$, is controlled by applying a gate
voltage. Modeling the charging energy on the dot by a Hubbard
on-site repulsion $U$, the system is described by a generalized
three-lead Anderson impurity model:
\begin{eqnarray}
{\cal H} =&& \sum_{\alpha =1}^{3} \sum_{k, \sigma}
          \left[ ( \epsilon_k + \mu_{\alpha} )
                   c^{\dagger}_{\alpha k \sigma} c_{\alpha k \sigma }
                 + t_{\alpha } \left(
                   c^{\dagger }_{\alpha k \sigma } d_{\sigma }
                   + h.c. \right)
          \right]
\nonumber\\
&& \;\;\;\;\;\;\;\;\;\;\;\;\;
             + E_D\sum_{\sigma} d^{\dagger}_{\sigma}d_{\sigma}
             + U d^{\dagger}_{\uparrow}d_{\uparrow}
                 d^{\dagger}_{\downarrow}d_{\downarrow} .
\end{eqnarray}
Here $c^{\dagger}_{\alpha k \sigma}$ creates an electron with wave
number $k$ and spin projection $\sigma$ in lead $\alpha$, while
$d^{\dagger}_{\sigma}$ creates a dot electron with spin
projection $\sigma$. In the following we assume that the
tunneling matrix elements and the energy level of the dot are
adjusted in such a way that a stable local moment is formed on
the dot. This corresponds to the condition
$\Gamma \equiv \sum_{\alpha=1}^{3} \Gamma_{\alpha} \ll -E_D, E_D+U$,
where $\Gamma_{\alpha} = \rho_{\alpha} \pi t_{\alpha}^2$
is the hybridization width associated with lead $\alpha$
($\rho_{\alpha}$ is the conduction-electron DOS at the Fermi
level of lead $\alpha$). We shall generally regard $\Gamma_3$
as much smaller than $\Gamma_1$ and $\Gamma_2$, though
comparable couplings will also be considered.

The main quantities of interest are the electrical currents
flowing from the dot to each of the three leads. The electrical
current flowing into lead $\alpha$ is given by the
nonequilibrium average of the current operator
$\hat{I}_{\alpha} = it_{\alpha} (e/\hbar) \sum_{k \sigma}
\{ c^{\dagger}_{\alpha k \sigma}d_{\sigma} - h.c. \}$
($-e$ is the electron charge). This average is conveniently
expressed via the lesser and retarded Keldysh Green functions
for the dot electrons: 
\begin{eqnarray}
&& G_{d\sigma}^<(\epsilon) =
       \int_{-\infty}^{\infty} \langle d^{\dagger}_{\sigma}
       d_{\sigma}(t) \rangle e^{i\epsilon t} dt , \\
&& G_{d\sigma}^r(\epsilon) = -i \int_{0}^{\infty}
       \langle \{ d_{\sigma}(t), d^{\dagger}_{\sigma} \}
       \rangle e^{i\epsilon t} dt .
\end{eqnarray}
Specifically, generalizing the expression of Meir and
Wingreen~\cite{WM92} from two to three leads one obtains
\begin{equation}
I_{\alpha} = \frac{4e}{h}\Gamma_{\alpha}\!\int_{-\infty}^{\infty}\!\!
             \left[ 2 \pi f(\epsilon\!-\!\mu_{\alpha})
	     A_{d}(\epsilon) -G_{d}^{<}(\epsilon) \right]
             \nu_{\alpha}(\epsilon\!-\!\mu_{\alpha}) d\epsilon ,
\label{I_alpha}
\end{equation}
where $A_{d}(\epsilon) = -\frac{1}{\pi}{\rm Im}\{G_{d}^{r}(\epsilon)\}$
is the dot spectral function (i.e., DOS);
$f(\epsilon)$ is the Fermi-Dirac distribution function; and
$\nu_{\alpha}(\epsilon) = \rho_{\alpha}(\epsilon)/\rho_{\alpha}$
is the reduced conduction-electron DOS in lead $\alpha$.
Here we use the notation by which the arguments of
$\rho_{\alpha}(\epsilon)$ and $\nu_{\alpha}(\epsilon)$ are
measured relative to the corresponding chemical potential,
$\mu_{\alpha}$. Note that we have restricted ourselves in
Eq.~(\ref{I_alpha}) to a zero magnetic field, in which case
all spin dependences drop.

The integration over energy in Eq.~(\ref{I_alpha}) contains
a natural cut-off, namely, the maximal chemical-potential
difference between the leads, broadened by the temperature.
To the extent that one can neglect the energy dependence of
$\nu_{\alpha}(\epsilon)$ on that scale, it is possible to
eliminate $G_{d}^{<}$ from the expression for $I_3$ using
the identity $I_3 = I_3(\Gamma_1+\Gamma_2)/\Gamma
- (I_1 + I_2)\Gamma_3/\Gamma$:
\begin{eqnarray}
I_{3} = \frac{4e\Gamma_3}{\hbar \Gamma}\!\int_{-\infty}^{\infty}\!\!
        A_{d}(\epsilon) && \left [
        (\Gamma_1 + \Gamma_2) f(\epsilon\!-\!\mu_{3})
	- \Gamma_1 f(\epsilon\!-\!\mu_{1}) \right.
\nonumber \\
        && \;\; \left.
	- \Gamma_2 f(\epsilon\!-\!\mu_{2}) \right ] d\epsilon .
\label{I_via_A}
\end{eqnarray}
Analogous expressions apply to each of $I_1$ and $I_2$.

\begin{figure}[tb]
\centerline{ \vbox{ \epsfxsize=80mm \epsfbox {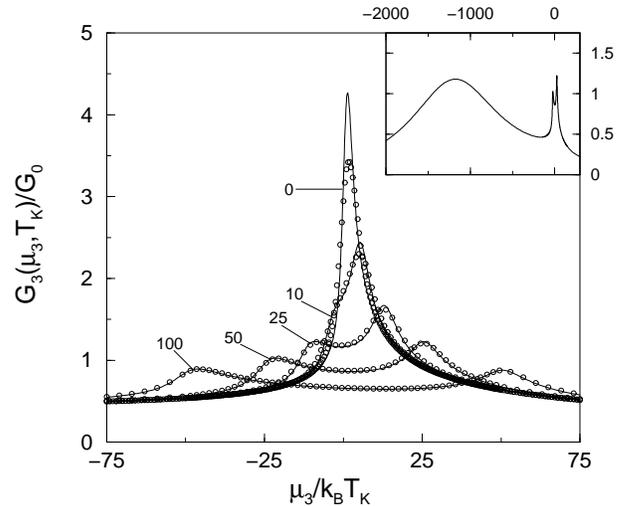}}
}
\vspace{10pt}
\caption{
     Comparison of the out-of-equilibrium dot DOS for a two-lead
     system (full lines) with the differential conductance
     $G_{3}(\mu_3)$ in the proposed three-lead setting (empty
     circles). All calculations were performed using the NCA.
     Here $\Gamma_1 = \Gamma_2 = D/30$
     (symmetric coupling), $E_d/D = -0.278$,
     $\Gamma_3/\Gamma_{1} = 0.01$, $T = T_K$, and
     $G_0 = 4e^2\Gamma_3(\Gamma_1 + \Gamma_2)/D\hbar \Gamma$. The
     two-lead dot DOS is scaled by the
     bandwidth $D$. The source-drain voltage bias,
     $\mu_2 - \mu_1 = eV_{sd}$, takes the values
     $eV_{sd}/k_B T_K = 0, 10, 25, 50,$ and $100$, as indicated
     by the attached labels. Up to thermal broadening, the scaled
     differential conductance coincides with the out-of-equilibrium
     dot DOS, in accordance with Eq.~(\ref{G_3_final}).
     Inset: An extended image of the dot DOS for a two-lead
     system with $T = T_K$ and $eV_{sd}/k_B T_K = 50$.}
\label{fig:fig2}
\end{figure}

Equation~(\ref{I_via_A}) generalizes the standard two-lead
expression for tunneling through an Anderson impurity~\cite{WM92}.
It retains the familiar form of an integral of a spectral
function times a weighted difference of Fermi functions.
If $A_d(\epsilon)$ were insensitive to variations in the
chemical potentials (as is the case for conventional noninteracting
systems), then the derivative of $I_3$ with respect to any
of the $\mu_{\alpha}$'s would have given the thermally
broadened dot DOS at $\mu_{\alpha}$. Clearly, this is not
the case for a quantum dot in the Kondo regime. Nevertheless,
for $\Gamma_3 \ll \Gamma_1 + \Gamma_2$ one can still
fix the chemical potentials $\mu_1$ and $\mu_2$ and differentiate
the current $I_3$ with respect to $\mu_{3}$, to obtain the
out-of-equilibrium two-lead dot DOS for a source-drain voltage
bias of $\mu_2 - \mu_1 = eV_{sd}$. To see this we note that,
for $\Gamma_3 \ll \Gamma_1 + \Gamma_2$, the impurity Green
functions only weakly depend on the coupling to the
third lead. Hence $A_{d}(\epsilon)$ and $G_{d}^{<}(\epsilon)$
are insensitive to variations in $\mu_3$, despite the
strong dependence on $\mu_1$ and $\mu_2$.
Fixing $\mu_1$ and $\mu_2$ and sweeping the bias $V_{31}$
(see Fig.~\ref{fig:fig1}), the differential conductance
$G_3(\mu_3) = dI_{3}/dV_{31}$ reduces then to
\begin{equation}
G_3(\mu_3) \simeq
	- \frac{4e^2 \Gamma_3}{\hbar \Gamma} (\Gamma_1 + \Gamma_2)
	\int_{-\infty}^{\infty}\!  A_{d}(\epsilon\!+\!\mu_3)
	\frac{\partial f(\epsilon)}{\partial \epsilon} d\epsilon .
\label{G_3_final}
\end{equation}
Hence, up to rescaling, $G_3(\mu_3)$ is just the thermally
broadened spectral function $A_{d}(\mu_3)$, which in effect
is the two-lead dot DOS for the chemical potentials $\mu_1$
and $\mu_2$. This means that the weakly coupled third lead
acts as a conventional tunneling probe for the dot DOS, even
though the dot itself is far from thermal equilibrium.

To test the above result, we have computed the
differential conductance $G_3(\mu_3)$ by direct numerical
differentiation of the current $I_3$ of Eq.~(\ref{I_alpha})
with respect to $V_{31}$, without resorting to
Eqs.~(\ref{I_via_A}) and (\ref{G_3_final}). Focusing for convenience
on the limit $U \to \infty$ (double occupancy is forbidden
on the dot), we evaluated the impurity Keldysh Green
functions using the noncrossing approximation (NCA)~\cite{Bickers87}.
The NCA is a self-consistent perturbation theory
about the atomic limit, which is known to provide a good
quantitative description of the temperature range
$T \geq T_K$. It has been extensively used to study dilute
magnetic alloys~\cite{Bickers87}, and has been successfully
applied to the out-of-equilibrium Kondo effect for both
single-channel~\cite{WM94,HS95,Langreth_et_al} and two-channel
scatterers~\cite{HKH94}. In the present context it has the
crucial advantage that it can be easily generalized to a
multi-lead setting. As a large-$N$ theory, though, the
NCA fails to recover Fermi-liquid behavior at low
temperatures~\cite{Bickers87}, and it overshoots the
unitary limit for $T < T_K$ in the $N = 2$,
nondegenerate case. To avoid these NCA pathologies,
we restrict attention hereafter to the range $T \geq T_K$.

Figure~\ref{fig:fig2} compares the the out-of-equilibrium
dot DOS for a symmetrically coupled two-lead system with
the differential conductance $G_{3}(\mu_3)$ in the proposed
three-lead setting. Here and throughout the paper we set
$(\mu_1 + \mu_2)/2$ as our reference energy, and use
a semi-circular conduction-electron density of states with
half-width $D$: $\nu_{\alpha}(\epsilon) = \sqrt{1 - (\epsilon/D)^2}$.
The impurity model parameters are $E_d/D=-0.278$ and
$\Gamma_1 = \Gamma_2 = D/30$, corresponding to a Kondo
temperature of
$k_B T_K/D = 2.5\!\times\!10^{-4}$~\cite{Comment_on_TK}.
The third-lead coupling, $\Gamma_3/\Gamma_1 = 0.01$, is
sufficiently weak as not to affect $T_K$.

Up to thermal broadening, the scaled differential conductance
of Fig.~\ref{fig:fig2} coincides with the out-of-equilibrium
two-lead dot DOS, in excellent agreement with
Eq.~(\ref{G_3_final}). For zero source-drain voltage bias,
$e V_{sd} = 0$, there is a single Kondo peak in $G_3(\mu_3)$,
corresponding to the thermally broadened Abrikosov-Suhl
resonance in $A_{d}(\epsilon)$. [Note that the height of
the resonance in $A_{d}(\epsilon)$ is somewhat lower than
$1/\pi \Gamma$, indicating that the Kondo effect is not yet
fully developed.] As the
source-drain bias is increased, a two-peak structure
gradually develops in $G_3(\mu_3)$. Two well-resolved peaks
are finally observed once $eV_{sd}$ exceeds
$k_B T_K$ and $k_B T$ by an order of magnitude or more.
For $eV_{sd} \gg k_B T, k_B T_K$, the effect of thermal
broadening is quite negligible, as the widths of the two
Kondo peaks are governed by the dissipative lifetime induced
by $V_{sd}$~\cite{WM94}, rather than by $k_B T$ or $k_B T_K$.
The asymmetric Kondo line shapes, particularly for small
to intermediate values of $e V_{sd}$, stem from the absence
of particle-hole symmetry in the infinite-$U$ Anderson
model.

\begin{figure}[tb]  
\centerline{ \vbox{ \epsfxsize=80mm \epsfbox {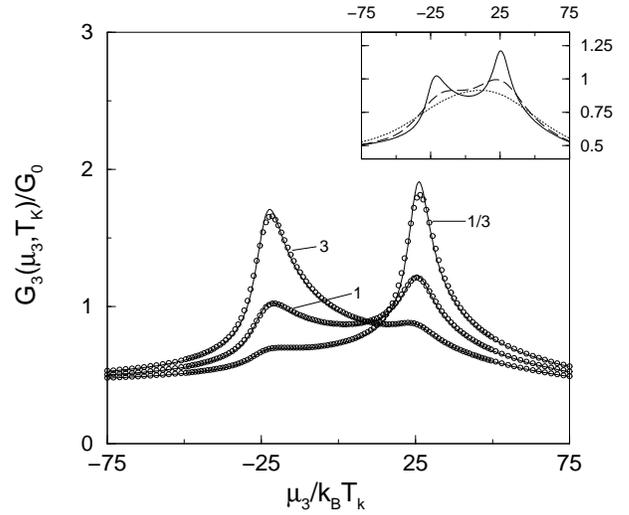}}
}
\vspace{10pt}
\caption{
     The out-of-equilibrium two-lead dot DOS (full lines) versus the
     three-lead differential conductance $G_{3}(\mu_3)$ (empty
     circles), for a fixed source-drain voltage bias of
     $eV_{sd}/k_B T_K = 50$, and different ratios of $\Gamma_1$
     to $\Gamma_2$. Here $(\Gamma_1 + \Gamma_2)/D = 0.067$ is
     kept fixed as to maintain the same Kondo temperature,
     $k_B T_K/D = 2.5 \times 10^{-4}$. The remaining parameters
     are $E_d/D = -0.278$,
     $\Gamma_3/(\Gamma_1 + \Gamma_2) = 0.005$, $T = T_K$,
     and $G_0 = 4e^2\Gamma_3(\Gamma_1 + \Gamma_2)/D\hbar \Gamma$.
     The ratio $\Gamma_1/\Gamma_2$ takes the values $1/3, 1,$
     and $3$, according to the labels attached. The effect of
     an asymmetry in $\Gamma_1$ and $\Gamma_2$ is to enhance
     (reduce) the peak at the chemical potential of the more
     strongly (weakly) coupled lead. Inset: Temperature dependence
     of $G_{3}(\mu_3)$, for $eV_{sd}/k_B T_K = 50$. Full,
     dashed and dotted lines correspond to $T/T_K = 1, 5,$
     and $10$, respectively. The effect of a temperature is
     to smear the peak structure of $G_{3}(\mu_3)$.}
\label{fig:fig3}
\end{figure} 	

Figure~\ref{fig:fig3} depicts the out-of-equilibrium two-lead dot
DOS versus $G_{3}(\mu_3)$, for $eV_{sd}/k_B T_K = 50$, $T = T_K$,
and different ratios of $\Gamma_1$ to $\Gamma_2$. The couplings
$(\Gamma_1 + \Gamma_2)$ and $\Gamma_3/(\Gamma_1 + \Gamma_2)$
are kept fixed as to maintain the same $T_K$. As before,
excellent agreement is obtained between $G_{3}(\mu_3)$ and
the two-lead dot DOS, in accordance with Eq.~(\ref{G_3_final}). The
effect of an asymmetry in the couplings to the source and drain
leads is to enhance the peak at the chemical potential of the more
strongly coupled lead, at the expense of the peak at the chemical
potential of the more weakly coupled lead. As seen in
Fig.~\ref{fig:fig3}, the effect is quite dramatic. Only a
shallow peak is left at the chemical potential of the more weakly
coupled lead when the ratio between the two couplings is three.
Accordingly, the dot DOS increasingly resembles that for sole
coupling to just a single lead. The inequivalent line shapes for
$\Gamma_1/\Gamma_2 =3$ and $\Gamma_2/\Gamma_1 =3$ are again the
result of the absence of particle-hole symmetry in our model.

\begin{figure}[tb]  
\centerline{ \vbox{ \epsfxsize=80mm \epsfbox {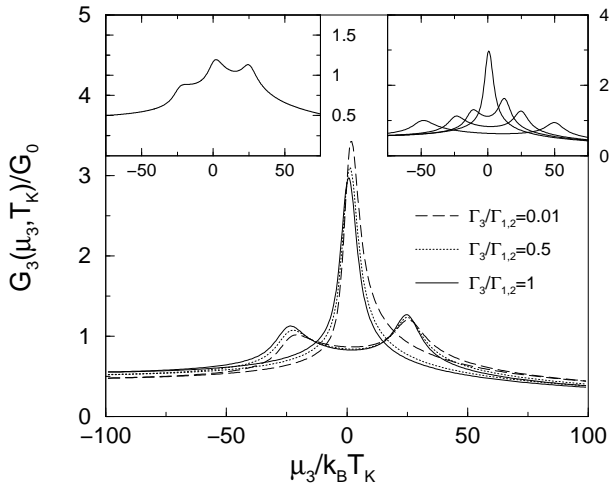}}
}
\vspace{10pt}
\caption{
     The three-lead differential conductance $G_{3}(\mu_3)$,
     for $\Gamma_1 = \Gamma_2$ and different ratios of
     $\Gamma_3$ to $\Gamma_1$.  The source-drain voltage
     bias, $eV_{sd}/k_B T_K$, is equal to zero for the
     three single-peak curves, and $50$ for the three
     split-peak curves. $\Gamma/D = 0.067$ is kept
     fixed in all curves as to maintain the same Kondo
     temperature. Here $E_d/D=-0.278$, $T = T_K$, and
     $G_0 = Dh\Gamma/8e^2\Gamma_3(\Gamma_1 + \Gamma_2)$.
     The overall peak structure of $G_{3}(\mu_3)$ is mostly
     unchanged when $\Gamma_3/\Gamma$ is increased. Right
     inset: The evolution of $G_{3}(\mu_3)$ with increasing
     source-drain voltage bias, for $\Gamma_1 = \Gamma_2 = \Gamma_3$
     and $T = T_K$. Here $eV_{sd}/k_B T_K$ equals $0, 25, 50,$
     and $100$. Left inset: The dot DOS, $DA_d(\epsilon)$, for
     $eV_{sd}/k_B T_K = 50$, $\mu_3/k_B T_K = 0$, and
     $\Gamma_1 = \Gamma_2 = \Gamma_3$. There are three
     separate peaks in the dot DOS, one peak at the
     chemical potential of each of the three leads.}
\label{fig:fig4}
\end{figure}    

The effect of a temperature is to smear the peak structure of
$G_{3}(\mu_3)$, as seen in the inset of Fig.~\ref{fig:fig3}.
This smearing stems both from the thermally broadened Kondo peaks
in the dot DOS, and from the convolution with the derivative of
the Fermi-Dirac function in Eq.~(\ref{G_3_final}). The two-peak
structure of the Kondo resonance remains visible, though, up to
$k_B T$ about an order of magnitude smaller than $eV_{sd}$ (see
also Fig.~\ref{fig:fig2}). For large source-drain bias, as in
Fig.~\ref{fig:fig3}, the two-peak structure thus extends up to
temperatures well above $T_K$.

Thus far our discussion has been restricted to weak coupling between
the dot and third lead. One may ask how does the qualitative
picture change once this coupling is no longer weak, especially as
the reasoning leading to Eq.~(\ref{G_3_final}) no longer holds.
In Fig.~\ref{fig:fig4} we have plotted $G_3(\mu_3)$ for
$\Gamma_1 = \Gamma_2$ and different ratios of $\Gamma_3$ to
$\Gamma_1$. Here $\Gamma$ is kept fixed as to maintain the same
$T_K$. Despite the notable changes in the dot DOS,
$G_3(\mu_3)$ remains mostly unchanged even for $\Gamma_3$ as
large as $\Gamma_{1}$. Indeed, while $A_{d}(\epsilon)$ acquires
a three-peak structure for $\Gamma_1 = \Gamma_2 = \Gamma_3$
(see Fig.~\ref{fig:fig4}, left inset), $G_3(\mu_3)$ remains closely
related to the two-lead dot DOS for the chemical potentials $\mu_1$
and $\mu_2$. This clearly shows the robustness of the proposed
method for measuring the splitting of the Kondo resonance.

A.S. is indebt to D. Goldhaber-Gordon, for inspiring
comments~\cite{Goldhaber-Gordon_Pecs}
during the NATO Workshop on ``Size-Dependent Magnetic Scattering,''
P\'ecs 2000. Discussions with O. Agam and V. Zevin are
gratefully acknowledged. This work was supported in part by the Centers
of Excellence Program of the Israel science foundation, founded
by The Israel Academy of Science and Humanities.


\begin{thebibliography}{9}
\bibitem{Comment_on_TK} We define $T_K$ as the temperature at which
        $R(T) = \left [ -\int_{-\infty}^{\infty} A^{-1}_{d}(\epsilon)
        \frac{\partial f(\epsilon)}{\partial\epsilon}d\epsilon
	\right ]^{-1}$ reduces to 50\% of its extrapolated
	zero-temperature limit. This definition of $T_K$ is common
	in dilute magnetic alloys, where $R(T)$ corresponds to
	the electrical resistivity due to scattering off
	the magnetic impurities. With $T_K$ so defined, the
	half-width of the Abrikosov-Suhl resonance in the
	NCA impurity spectral function is about $1.7k_B T_K$.
\end{thebibliography}
\end{document}